# Voiced speech as secondary response of a self-consistent fundamental drive

Friedhelm R. Drepper

Forschungszentrum Jülich GmbH, 52425 Jülich, Germany, f.drepper@fz-juelich.de

Voiced segments of speech are assumed to be composed of non-stationary acoustic objects which can be described as stationary response of a non-stationary fundamental drive (FD) process and which are furthermore suited to reconstruct the hidden FD by using a voice adapted (self-consistent) part-tone decomposition of the speech signal. The universality and robustness of human pitch perception encourages the reconstruction of a band-limited FD in the frequency range of the pitch. The self-consistent decomposition of voiced continuants generates several part-tones which can be confirmed to be topologically equivalent to corresponding acoustic modes of the excitation on the transmitter side. As topologically equivalent image of a glottal master oscillator, the self-consistent FD is suited to serve as low frequency part of the basic time-scale separation of auditive perception and to describe the broadband voiced excitation as entrained (synchronized) and/or modulated primary response. Being guided by the acoustic correlates of pitch and loudness perception, the time-scale separation avoids the conventional assumption of stationary excitation and represents the basic decoding step of an advanced precision transmission protocol of self-consistent (voiced) acoustic objects. The present study is focussed on the adaptation of the trajectories (contours) of the centre filter frequency of the part-tones to the chirp of the glottal master oscillator.

*Keywords:* signal analysis, speech recognition, voiced continuants, part-tone phases, fundamental drive, cascaded response, generalized synchronization

## 1. Introduction

For decreasing signal to noise ratio, human speech perception shows an outstandingly lower increase of the error rate of word recognition, when compared to automatic speech recognition (ASR) [1, 2]. The higher robustness of human acoustic perception has been demonstrated in many different situations of speech communication including context free presentation and the cocktail party. In the latter situation the use of frequency modulated voiced speech turns out to generate more robust communication than the use of unmodulated voiced speech [3] and in particular than the use of whispered speech. When combined with the higher sensitivity of ASR and of whispered speech to the distance and direction of the acoustic communication, these empirical findings open an unconventional perspective on the reasons of the well known problems of ASR. These problems may be caused to a lesser extent by the obvious imperfection of the linguistic or statistical language models and to a larger extent by an insufficiently differentiated description of the aero-acoustics of voiced speech.

In spite of the undisputedly high degree of non-stationarity of speech signals, the present day determination of acoustic feature vectors of ASR is based on the assumption that speech production can be described as a linear time invariant (LTI) system (on the time scale of about 20 ms). The wide sense stationarity of an LTI–system is typically used either as prerequisite for the consistent estimation of Fourier spectra or of autoregressive (all pole) models [2, 4, 5]. In the latter case it is common practice to introduce a drive-response (input-output or source-filter) model, which restricts the stationary autoregressive description to the resonance properties of the vocal tract. Linear autoregressive models are suited to describe transients with varying decay rates in different frequency ranges including relatively long (resonant formant) transients. The average decay rates (Liapunov exponents) of such transients are known to represent topological invariants [6] (which can be assumed to be invariant under changes of the geometry of the acoustic transmission) and important cues for the distinction of vowels [2, 4, 5]. However, the conventional LTI system approach turns out to be problematic in the case of voiced phones. The vocal tract filter should not be assumed to be time invariant [2, 4, 5] and the source not to be generated by an autonomous *linear* dynamical system [7-10].

Low dimensional autonomous *nonlinear* dynamical systems have been introduced to describe newborn infant cries and dysphonic adult voices [9] and have also been found to bring additional accuracy to models of normophonic voices [10]. However, there is empirical evidence that the complex neural control of the vocal fold dynamics leading to shimmer, jitter and vocal tremor [11] impedes or precludes a low dimensional *autonomous* deterministic description of the phonation process. Being hopelessly irregular from the point of view of acoustics, the time evolution of pitch and loudness (intonation and prosody) can partially be given phonological interpretation [32]. The connection to linguistics and paralinguistics (emotions) invalidates or challenges a *stationary* stochastic process description of the voice source.

A more differentiated and physiologically plausible phenomenological description of the aero-acoustics of voiced speech can be achieved by introducing an additional drive-response step, which describes the highly complex wideband acoustic source as stationary (primary) response of a *non-stationary*, band-limited fundamental drive process in the frequency range of the pitch [12-16]. The importance, generality and precision of the acoustic percept of pitch can be taken as a first hint that the hidden fundamental drive (FD) can directly be extracted from the speech signal. This leads to a two-level cascaded drive-response model (DR model) of voiced speech production which describes the speech signal as secondary response of a hidden FD. The two levels of the response cannot only be interpreted (more or less erroneously) as source and vocal tract filter output but can also be used with advantage to introduce two complementary types of simplification of the cascaded response dynamics.

In case of the secondary response it is common practice to simplify the vocal tract resonances by assuming time invariant stable *linear* response dynamics (resulting from an all pole filter) with a fixed point attractor. (An attractor is an invariant set of states which homes the asymptotic long time behaviour of the dynamics. Stable linear dynamical systems have a single trivial point attractor with dimension $d = 0$, the origin of state space). In case of the primary response there is no doubt that voiced continuants (sustainable phones with active phonation) are generated by *nonlinear* dynamics with attractors of dimension $d > 0$ [9-11]. Complementary to the long transients of a typical (vowel type) secondary response, the primary response is assumed to result from strongly dissipative nonlinear dynamics [6], which generates predominantly short transients. Such dynamics can be simplified drastically by restricting the dynamics to the asymptotic invariant set (which neglects the transient behaviour).

Invariant sets (attractors) with dimension $d > 0$ of *autonomous* (nonlinear) deterministic dynamical systems are known to represent either continuous manifolds (limit cycles or tori) or fractal sets (homing chaotic dynamics) [6], whereas *uni-directionally coupled* (DR) systems with dissipative responses are known to have invariant sets which are subsets of continuous synchronization manifolds (lines or surfaces) in the combined state space of drive and response [17, 18]. Being constrained to a continuous synchronization manifold, the (primary) response can be expressed by a continuous coupling function which describes the momentary state of the response by a unique function of a response related state of the (fundamental) drive. Synchronization or phase locking is known to be a generic property of nonlinearly coupled DR dynamics [6]. As a rather general form of synchronization of band-limited oscillators with different frequencies, phase locking (synchronization of phases) can be detected by choosing an oscillator description in terms of amplitude and phase variables and by restricting the synchronization analysis to the phases of the oscillators [19].

As will be explained in more detail, the distinction between the acoustic source and the FD opens the option to reconstruct a coherent hidden drive for a complete voiced speech segment. The latter feature can be used to reveal additional features of voiced speech, which are invariant (robust) under changes of the acoustic communication channel and to separate the phonetically relevant fast dynamics from intonation and prosody without invoking the assumption of stationary excitation.

The idea that the higher frequency acoustic modes of voiced speech and music as well as the perception of their pitch are causally connected to a single acoustic mode in the frequency range of the pitch (*son fundamentale* or fundamental bass), can be traced back to Rameau [20]. However, Seebeck [21] could show that (virtual) pitch perception does not rely on a fundamental acoustic mode which is part of the heard signal. (In the meantime Fourier had invented an efficient method to separate *periodic* modes with different frequencies.) Seebeck was also the first to use a time periodic "impulse function" which can be seen as a predecessor of the above coupling function, to describe the sound of sirens. The latter sound sources might be interpreted as highly simplified phonation systems. More recently a "wave shaper" function of an *aperiodic* synthetic drive has been used by Schoentgen to describe [22] and synthesize vowels with normal and disordered jitter and micro-tremor [23]. In preliminary studies the principle feasibility of a reconstruction of the FD from a voiced speech signal as well as from a simultaneously recorded electro-glottogramme has been demonstrated and compared [12-14]. However, the extraction of the FD had been based on a subband decomposition with critical (audiological) bandwidths, which relied on short term *stationary* bandpass filters (with time independent centre filter frequencies within the current window of analysis). Signal decompositions with a similar deficiency have also been used in several other studies [24-27, 33].

The present study describes a phonation-physiologically, aero-acoustically and psychoacoustically plausible decoder, which is suited to reconstruct the hidden FD of a *non-stationary* voiced speech segment and to confirm the topological equivalence of the FD to a glottal master oscillator of the voice source. This is achieved by replacing the assumption of a short term stationary (periodic) voice source by the more general assumption of synchronized dynamics of a time dependent subset of part-tone phases of the speech signal. Section 2 is focussed on the plausibility of the central synchronization hypothesis and outlines the reconstruction of the fundamental phase. Section 3 outlines the reconstruction of the fundamental amplitude as well as of the primary response and proposes two additional topological invariants of the latter as cues for speaker and/or phoneme recognition. Section 4 describes a self-consistent (voice adapted) part-tone decomposition of the speech signal which is suited for a high precision reconstruction of the fundamental phase (velocity). Section 5 uses an elementary chirped phonation process to illustrate a self-consistent part-tone decomposition and is followed by a discussion and a conclusion.

## 2. AUTO-PHASE-LOCKING OF THE VOICED EXCITATION

The phenomenon of synchronization of coupled *periodic* oscillators is known since the time of C. Huygens. More recently, synchronization between *aperiodic* oscillators has been analysed, in particular the cases of unidirectional coupling to stochastic [17] or deterministic chaotic [18] drives. In these cases the synchronization manifold describes the momentary state of the response by a unique continuous (coupling) function of the simultaneous state of the drive. The next more complex cases of synchronization in DR systems are characterized by a finite number of coupling functions which define a set-valued mapping, which generates different responses depending on the most recent past of the drive process. The latter feature characterizes the so called voice type. (Diplophonia e.g. describes the case of a bi-valued mapping or of a period doubling of entrained oscillatory degrees of freedom of the glottal dynamics [9, 11, 15].)

Continuous synchronization manifolds are not limited to oscillators with exclusively unidirectional coupling. The notion of a DR system will therefore be used in the following also for dominantly unidirectionally coupled systems. Due to the large density ratio of about 1000:1 between tissue and air, the convective dynamics and the resulting acoustic modes (fast oscillatory degrees of freedom of the aero-acoustic dynamics) in the larynx adjust nearly instantaneously to the time varying boundary conditions of the air-duct, which are given or dominantly influenced by the comparatively slow motion of the vocal folds [9, 11]. In spite of a potential back-coupling of acoustic modes to the dynamics of the vocal folds, the coupling of the acoustic modes of the larynx to the vocal folds can be assumed to have a dominant direction of interaction which is implied by assuming DR dynamics.

A particularly simple type of synchronization occurs in case of the near symmetrically coupled dynamics of the two vocal folds as is typical for non-pathological phonation [9, 11]. The synchronization manifold of a symmetrically coupled oscillator system can be described by a coupling function representing a conjugation (continuous one to one mapping with a continuous inverse). This means that the two oscillators become topologically equivalent [6], i.e. they behave like a single oscillator. Except in the severely pathological case of so called biphonation (incommensurate frequencies of the two vocal folds) [9], a glottal master oscillator with a single independent degree of freedom (e.g. the area of the glottis or its width) can be assumed to exist, which synchronizes many of the other oscillatory degrees of freedom of the larynx.

The time profile of the glottal pulses is well suited to excite a broad spectrum of *acoustic* modes with frequencies reaching far above the one of the glottal pulses [2, 4, 5]. Due to the *nonlinear* coupling of the acoustic modes of the larynx to the glottal master oscillator the higher frequency modes show a typical ($n:1$) synchronization (phase-locking) [6, 9, 18]. The idea of Rameau that a single fundamental acoustic mode drives its harmonic overtones [20] can thus be extended to the idea that a glottal master oscillator with a single degree of freedom synchronizes the acoustic modes of the larynx (primary response) and synchronizes or drives the acoustic modes of the vocal tract (secondary response). This is indicated in the upper part of figure 1.

For not too high harmonic order $n$ the typical ($n:1$) phase-locking between voiced acoustic modes of the vocal tract and the glottal master oscillator becomes potentially visible (audible) as a mutual ($n:n'$) phase-locking of appropriately chosen part-tone pairs of the speech signal [12-15]. The mutual phase locking can, however, only be detected, if the set of (preferentially complex) bandpass filters being used to separate the part-tones and to generate their phases is adapted with high accuracy to the instantaneous frequency of the glottal master oscillator. The need of accuracy of the adaptation increases with the maximum harmonic order $n$ or $n'$. In view of the ubiquitous micro-tremor and jitter of the glottal oscillator, the centre frequencies of the part-tones should preferentially be chosen as time dependent, even within a single analysis window [16]. (That is why the notion "part-tone" is preferred to the more common notion "subband" [25-27].)

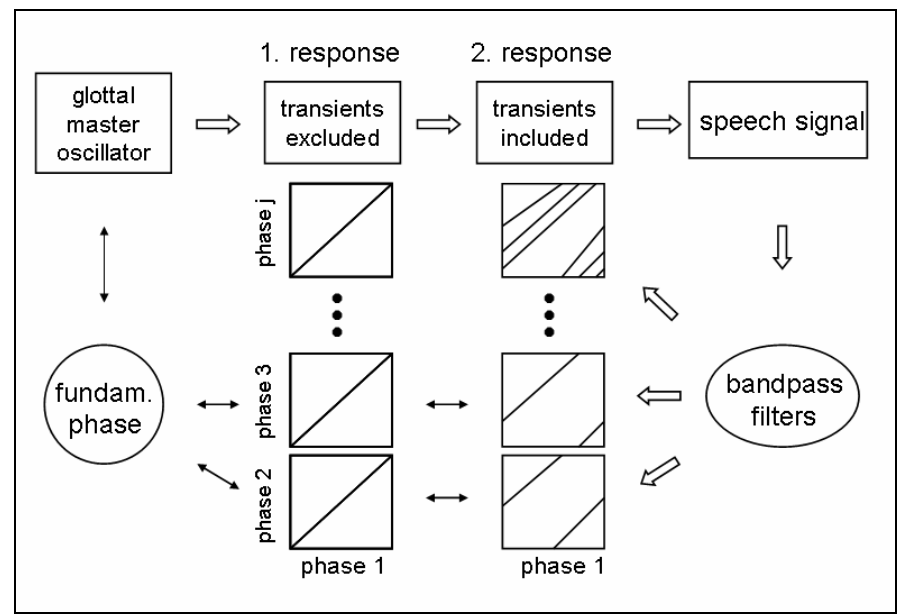

**Figure 1**: Causal connections between the glottal master oscillator and two cascaded responses which can vaguely be interpreted as voice source and vocal tract output. Some part-tone phases of the speech signal show invertible relations to the phase of the fundamental drive (FD) which can be confirmed to be topologically equivalent to the glottal master oscillator. Examples of invertible phase relations are given in figure 3.

The detection of a strict ($n:n'$) synchronization of the phases of *a priori* independent part-tone pairs (with non-overlapping spectral bands) represents a phenomenon, which has a low probability to happen by chance, in particular, when the higher harmonic order $n$ has a large value. For such part-tone pairs it can therefore be assumed that there exists an uninterrupted causal link between those part-tones, including the only plausible case of two uninterrupted causal links to a common drive, which can be identified as the glottal master oscillator (figure 1). Since the ($n:n'$) phase-locking with $n \neq n'$ is generated by the nonlinear coupling to the glottal oscillator, a stable synchronization of *a priori* independent part-tone phases can be taken as a confirmation of topological equivalence between these part-tones and respective acoustic modes in the vocal tract of the transmitter [12]. One of the part-tone phases (e.g. phase 1) can be used as a reference to summarize the larger number of pairs of mutually synchronized part-tone phases. The reduced set of phase relations is indicated in the lower part of figure 1.

The glottal master oscillator being part of the transmitter side, its hidden instantaneous frequency has to be replaced by the one of a topologically equivalent image on the receiver side (being called fundamental drive or FD). As an important property of non-stationary voiced acoustic objects it turns out that the phase velocity of the FD of voiced continuants can be reconstructed iteratively (self-consistently), when combined with the self-consistent choice of the bandpass filters. The main idea is that invertible phase relations can be used to trace back the causal link to the common FD (figure 1).

As has been anticipated more or less explicitly by Rameau [20], Seebeck [21] and Terhardt [26, 27], the perception of virtual pitch can be interpreted as an instrument of basic causality analysis for sufficiently simple acoustic DR systems with a hidden drive and an observable response. For stationary voiced sounds Terhardt has described the acoustic correlate of (nominal) virtual pitch as the frequency of maximal subharmonic coincidence, i.e. of the coincidence of an extended set of part-tone frequencies being each normalized by an appropriate set of harmonic orders. In the more general (non-stationary) case the frequency of maximal subharmonic coincidence has to be replaced by the largest cluster of trajectories (contours) of appropriately normalized part-tone phase velocities. The largest cluster of subharmonic coincidence defines a set of part-tones which can be grouped together to form a multi part-tone voiced acoustic object being characterized by a related set of phase normalization factors (winding numbers) (see below). Each of the normalized part-tone phase velocity contours (or a suitably defined centroid of the largest cluster) can be taken as phase velocity contour of a topologically equivalent image of the non-stationary glottal master oscillator and thus as velocity of the fundamental phase (figure 1).

The detection of an ($n:n'$) synchronization of part-tone phases is thus suited to confirm ("online" on the receiver side) that voiced continuants are characterized by a subset of acoustic modes of the vocal tract which are synchronized by the glottal master oscillator. For these phonemes there exists a potentially non contiguous frequency range, for which the initial ($n:1$) synchronization of the acoustic modes is not disturbed during the passage through the vocal tract, in particular not by the vocal tract resonances (formants) or an additional constriction of the air flow. The "transparent" frequency range of the vocal tract (not to be confused with the presumably empty frequency range where no phase shifts occur) can be expected to belong to the range of phase sensitive human acoustic perception, i.e. to the range of the first three or four formants (below 4 kHz).

As a further remarkable feature of the unexpectedly subtle voice transmission protocol being partially uncovered in this study, the reconstruction of the phase velocity of the FD can typically be achieved for considerably larger speech intervals than the one of the confirmation of topological equivalence between part-tones and respective acoustic modes of the vocal tract. Whereas the piecewise detection of subharmonic coincidence clusters is preferentially performed in a time window of about 30 – 40 ms, the reconstruction of a coherent (continuous) fundamental phase can hypothetically be achieved for time segments, which correspond to an uninterrupted voiced segment of a syllable, i.e. for a time span of typically more than 100 ms [13, 15]. For voiced segments, which are composed of different phonemes, it is typical that the set of part-tones being used for the piecewise reconstruction of the fundamental phase velocity contour changes in succeeding analysis windows.

According to a comparatively strict phonological rule [32], uninterrupted voiced segments of speech contain at least one vowel nucleus or another sustainable voiced nucleus with a comparable transparency of the glottal closure event (e.g. /l/ like in "title"). This nucleus can be used to gauge the (wrapped up) fundamental phase with respect to the glottal closure event and can thus be used to remove the arbitrariness of the initial phase of the FD of a voiced segment.

### 3. RECONSTRUCTION OF THE PRIMARY RESPONSE

In case of the reconstruction of the amplitude of the FD it is not far fetched to identify loudness perception as the complementary instrument of causality analysis. From many psychoacoustic experiments it is known that loudness perception represents an even more universal and unique acoustic percept than virtual pitch. It is therefore tempting to relate the instantaneous fundamental amplitude to a building block of the acoustic correlate of loudness perception. The latter one is usually described as an aggregate of contributions from a range of audiological (critical) subbands and from a range of observation times. The empirically well established first aggregation step combines the elementary contributions from the different subbands. The elementary contributions are based on instantaneous subband amplitudes $a_{i,t}$ and the aggregation is known to include a characteristic sublinear scaling (compression) of these amplitudes [28, 29]. The sublinear scaling is usually expressed by a power law with an exponent in the range of $v = 0.33$. It is also known that the first aggregation step includes a fairly complex subband specific weighting [28]. In a rough approximation the subband weights can be chosen as amplitude independent, e.g. as proportional to frequency dependent inverse hearing thresholds [28]. The latter ones are known to have a minimum in the range from 1000 to 3000 Hz.

Since the spectrum of typical voiced tonal signals shows a pronounced downhill slope towards higher frequencies, the strong compression of the amplitudes has the effect of a down-weighting of the lower frequency amplitudes. In combination with the frequency dependence of the subband weights this leads to a substantial down-weighting of the subbands below 1 KHz. The latter subbands are the potential candidates to experience group delays which are substantial fractions of the fundamental (glottal) cycle length due to vocal tract resonances. Thus the suppression of the lower frequency subbands supports the interpretation of loudness perception as the complementary instrument of basic causality analysis.

The described building block of loudness perception is assumed to be directly related to the high time resolution version $a_t$ of the fundamental amplitude $A_t$. The relation is chosen as the corresponding inverse power low, such that instantaneous amplitude $a_t$ represents a linear homogenous function of the set of subband amplitudes $a_{i,t}$.

$$a_t = \left( \sum_{j=1}^{N} (w_j a_{j,t})^v \right)^{\frac{1}{v}} \quad \text{with} \quad \sum_{j=1}^{N} w_j^v = 1 \ . \quad (1)$$

In an even more rough approximation the subband weights $w_j$ are chosen as proportional to the harmonic order of the subbands [5, 28]. For reasons of unification it is assumed that the set of subbands $j = 1, ..., N$ is chosen to include the set of self-consistent part-tones being used for the reconstruction of the fundamental phase. Due to the suppression of the lower frequency part-tones, the instantaneous amplitude $a_t$ is well suited to gauge the wrapped up fundamental phase with respect to the glottal closure event (of vowels). The time scale separation leading to the desired low time resolution version $A_t$ of the fundamental amplitude is performed in accordance to the one being used for the fundamental phase (see below).

The indicated reconstruction steps result in a unique fundamental amplitude $A_t$ and a fundamental phase $\psi_t$, which is uniquely defined within voiced segments of speech. As has been explained above, non-pathological speech is characterized by the fact that the action of the glottal master oscillator on the acoustic modes of the vocal tract can be described by a single independent oscillatory degree of freedom. The two variables $(A_t, \psi_t)$ can therefore be interpreted as a complete set of *response related state variables* of a FD, which is suited to reveal a set of self-consistent part-tones of a non-stationary voiced acoustic object and to describe the auto-phase-locking of its excitation as synchronized primary response. It is convenient to express the state of the fundamental drive as complex number (analytic signal) $A_t \exp(i\psi_t)$.

Thus the initially cited contradiction between the concepts of Rameau and Seebeck can now be reconciled by replacing Rameau's *son fundamentale* by the described FD. Being of a more abstract nature and in need of a confirmation of its existence, the FD of a *multi-part-tone* voiced acoustic object cannot be reconstructed from the output of a single bandpass filter alone. Furthermore its existence does not rely on the presence of an acoustic mode in the frequency range of the pitch. Both features support the role of the instantaneous fundamental phase velocity as acoustic correlate of *virtual* pitch perception.

Once the FD is reconstructed and confirmed, the non-stationary self-consistent (voiced) acoustic object, which is assumed to approximate the speech signal of a single analysis window (of about 30-40 ms), can be reconstructed as a deterministic or conditional stochastic response with a stationary coupling to the non-stationary FD [14-16]. The assumption of a two-step cascaded response brings the reconstruction into contact with the conventional source-filter model of voiced speech, i.e. the primary response is interpreted as the classical voice source. As alternative to the conventional definition as inversely filtered speech signal, the broadband voiced excitation is defined as primary response of the band-limited FD. In accordance to the definition of the fundamental amplitude (given in equation 1), the dependence of the primary response on the (low time resolution) fundamental amplitude $A_t$ is assumed to be a linear one [14-16]. As a basic option the primary response is described as sum of a strictly synchronized deterministic response of the form $A_t G(\psi_t)$ with the fundamental phase dependent coupling function $G(\psi)$ and of a stochastic response $A_t \sigma(\psi_t) \xi_t$ which is modulated by the product of the fundamental amplitude $A_t$ and a potentially fundamental phase dependent modulation function $\sigma(\psi)$. This leads to the following window specific primary response

$$E_t = A_t (G(\psi_t) + \sigma(\psi_t)\xi_t). \qquad (2)$$

The stationary linear noise process $\xi_t$ is assumed to have a short range autocorrelation (see below). Preferentially it is chosen as a white noise process with a Gaussian distribution. For a so called modal (normal) voice the deterministic skeleton $A_t G(\psi_t)$ is expected to dominate and for a breathy voice source the stochastic process part gets a comparable amplitude. Due to the infinite multiplicity of the mathematical definition of phases, all functions of phases, which express physically unique properties, have to be periodic. Therefore coupling function $G(\psi)$ can be approximated by a finite Fourier series, which makes use of the appropriate periodicity [12-16]. The period length of the phase functions does not necessarily have to be $2\pi$. In case of a diplophonic voice type the basic period may be chosen as an integer multiple $2\pi m$ with the subharmonic period number $m > 1$ [14-16]. In case of part-tones with a harmonic number $n > 1$ the synchronization is described by a multi-modal coupling function, which describes an $(n:m)$ phase locking to the FD (with a so called winding number $n/m$). The phase locking of the higher frequency acoustic modes of the voiced excitation is, however, partially destroyed by resonances (long transients) of the vocal tract. It is common practice to describe the latter effect by linear autoregressive filters [2, 4, 5].

To reconstruct voiced phones, primary excitation (2) is taken as input to a phone specific *linear* autoregressive filter which represents the secondary response of the cascaded DR model [14, 16]. The secondary response generates the transients which have deliberately been excluded from both parts of the primary response in equation (2). For some of the phones and/or for some parts of the frequency range, the secondary response might be interpreted as a vocal tract filter.

Due to their (more or less) direct relation to the vocal tract shape [2, 4, 5], the autoregressive parameters of the secondary response represent important cues for phoneme recognition. For a non-stationary FD, conditionally estimated autoregressive parameters are more closely related to the vocal tract shape than unconditionally estimated ones. In contrast to the conventional analysis of the resonance properties of the vocal tract [2, 4, 5], which is based on a phone and prosody unspecific premphasis (of the higher frequencies) of the speech signal, advanced precision autoregressive parameters have to be estimated simultaneously with the coefficients of the Fourier series expansion of $G(\psi)$. For the Gaussian white noise alternative of equation (2) (and a missing or given phase dependence of the modulation function $\sigma(\psi)$), the simultaneous estimation of all parameters of the cascaded response can in principle be achieved by multiple linear regression [14-16]. However special care has to be taken to avoid underdetermined parameters (due to missing orthogonality of the basis vectors) by using *a priory* knowledge about the pulse shape of the glottal excitation. A more detailed description of the simultaneous parameter estimation is outside the scope of the present study. The two-level cascaded response model is expected to be well suited for the transformation (pitch or vocal tract length manipulation) and high quality synthesis of voiced speech.

The description of the voiced excitation according to equation (2) generalizes the well known time periodic skeleton and unmodulated linear noise process of an LTI system generated excitation [2, 4, 5]. Whereas the LTI system has to be based on the assumption of a frequency gap, which separates the phonetically relevant fast dynamics from lower frequency micro-tremor and prosody, the time scale separation being based on a self-consistent FD and primary response (2) does no longer rely on such an assumption.

The coherent reconstruction of the fundamental phase turns out to be useful not only for the reconstruction and distinction of vowels but also of sustainable voiced consonants. For the acoustic correlates of the latter phonemes the simple source-filter interpretation in terms of a plane wave source in an unbranched vocal tract looses its validity and the nonlinearity of the aero-acoustic dynamics becomes more apparent. Nasals are characterized by a sudden apparent time shift of the glottal closure event (due to a sudden increase of the group delay of the acoustic transmission path from the glottis to the receiver) [14, 15]. Voiced fricatives (as the English /z/ like in "zoom") are characterized by an acoustic source in the vicinity of a second constriction of the vocal tract, which generates an intermittently turbulent jet. The conversion of the kinetic energy of this pulsatile jet into acoustic energy (e.g. at the edge of the teeth) happens with a characteristic time delay, which results from the comparatively slow subsonic convection speed of the relevant jet [7, 8]. The phoneme specific relative time shift of the second acoustic source (relative to the glottal closure event) should be interpreted as further example of a topological invariant which is comparatively insensitive to disturbances of the acoustic communication channel. To describe the second acoustic source of voiced fricatives, primary response (2) should be extended by a second additional deterministic coupling function $G_{II}(\psi_{t-\tau})$ with a phone specific time shift $\tau$.

Being of a predominantly Lamarckian type, the (cultural) evolution of speech can be assumed to have proceeded comparatively fast. The present adaptation of speech production to the abilities of the human auditive pathway can therefore be assumed to be so perfect, that the study of human (and mammalian) acoustic perception can be used to infer important properties of the aero-acoustics of speech production and vice versa. The astonishing ability of human acoustic perception to recognize a large number of different speakers and/or phonemes under varying geometries of the acoustic transmission (and beginning at a comparatively early stage of the ontogenetic evolution) cannot easily be explained without assuming that voiced speech is characterized by a rich spectrum of qualitatively different types of response dynamics which can be classified by a comparatively low number of cues. Furthermore these cues are expected to represent topological invariants of the acoustic dynamics i.e. quantities which are invariant under topological transformations (conjugations) of the state space, in particular integer valued invariants like dimensions and periodicities of synchronization manifolds. As described above, such ideal properties of voiced speech are in full agreement to the dissipation and nonlinearity of the speech production dynamics.

It is therefore consequent to hypothesize that topological invariants of the described cascaded DR dynamics of voiced speech (periodicities of part-tones, Liapunov exponents, time shifts and relative voice onset times) are collectively used by human acoustic perception as cues to distinguish speakers and phones. This psychoacoustic hypothesis is supported by the fact that the distinction of vowels is well known to be based on Liapunov exponents of the secondary response dynamics and the distinction of voiced vs. unvoiced stop consonants (like /b/ and /p/) by different voice onset times. In particular it is plausible to identify the potentially numerous combinations of (*n:m*) synchronization of different part-tone phases as examples of a qualitatively rich (speaker specific) behaviour of the primary response, complementing the well known qualitatively rich (phoneme specific) behaviour of the seconddary response. As has been pointed out by Terhardt [26], the focussing of *acoustic* perception on qualitative (topological) properties of the output of adaptively chosen bandpass filters parallels the gestalt perception of *optical* signal streams.

The coherent reconstruction of the fundamental phase of an extended voiced speech segment and the detection of higher order phase locking of its part-tones requires an extremely precise reconstruction of the phase velocity of the glottal master oscillator. This underlines the importance of a precise adaptation of the centre filter frequencies of the part-tone decomposition to the respective signal.

### 4. VOICE ADAPTED PART-TONES

Many analysis methods based on time-frequency energy distributions like short time Fourier analysis or wavelet analysis are based on complete, orthogonal decomposition of the (speech) signal into elementary components. In case of wavelet analysis the elementary components are chosen as near optimal time-frequency atoms, which are each characterized by a reference time $t_0$ and angular frequency $\Omega_0$. As a characteristic feature of wavelet analysis, the time-frequency atoms are chosen on different frequency scales. Time-frequency atoms are wave packets which are optimized to describe simultaneously event (particle) and wave type properties of non-stationary wave processes [30]. Their most general form can be written as (real part of) a second order logarithmic expansion of an analytic signal around the reference time $t_0$ resulting in a complex Gaussian of the form

$$S_G(t) \approx \exp(-\frac{(t-t_0)^2}{2\sigma^2} + i\Omega_0(t-t_0)(1+c/2(t-t_0))) \quad (3a)$$

Contrary to the conventional one [30], this parametric set of time-frequency atoms is characterized by a *quadratic* trend phase or a linear trend phase velocity (angular frequency)

$$\omega_{0,t} = \Omega_0(1+c(t-t_0)) \quad (3b)$$

with relative chirp rate $c$. In view of their neglect of the chirp parameter, the time-frequency atoms of short time Fourier analysis and wavelet analysis are preferentially aimed at LTI systems (with a time periodic deterministic skeleton). In contrast to the latter approaches the present one is aimed at non-stationary acoustic objects which represent a superposition of time frequency atoms with chirped angular frequencies. The general aim, however, is not a complete and orthogonal decomposition of the speech signal into time frequency atoms, but a decomposition into part-tones which result from a superposition of appropriately chosen time frequency atoms such that these part-tones can be interpreted as topologically equivalent images of plausible underlying acoustic objects. Like in auditory scene analysis, the *a priori* knowledge about the behaviour of the acoustic objects is used to remove a potential under-determinedness of the unknown acoustic object parameters which results from a potential non-orthogonality of the decomposition. As we will see in more detail, the additional chirp parameter of the time frequency atoms can be used to compose part-tones with a time evolution which is consistent with the one of their bandpass filters. Self-consistently reconstructed part-tones of voiced speech segments are expected to allow an interpretation in terms of the described qualitatively rich auto-phase locking phenomena. In case of the characteristic click type event of a stop consonant a single time frequency atom is potentially suited to describe this event.

In view of the large range of relevant time scales in speech communication, a time-scale separation is highly desirable. As is well known, pitch and loudness perception introduce a dual basic time-scale separation which separates the phonetically relevant time scales from all longer ones. Due to its coherence (continuity) within voiced segments of speech, the fundamental phase $\psi_t$ of the preceding section is an ideal candidate to be used for the long time scale (low frequency resolution) description. A second good candidate is the fundamental amplitude $A_t$. To achieve the appropriate smoothing of the variables of the long time scale description, the fundamental amplitude $A_t$ of the current window of analysis can e.g. be obtained by fitting a quadratic polynomial or a Gaussian to the time evolution of the instantaneous fundamental amplitude $a_t$ of equation (1) under the constraint of continuity and smoothness at the transition to the preceding analysis window. The smoothing of $\psi_t$ will be explained below.

For real time application it is unavoidable to use causal bandpass filters for the generation of the part-tones. The present study uses an all pole approximation of complex $\Gamma$-tone bandpass filters [31] with approximately gamma-distribution like amplitudes of the impulse response. For sufficiently high autoregressive order, the $\Gamma$-function like amplitude distribution guarantees a near optimal time-frequency atom property of the impulse responses. That is why an autoregressive order ($\Gamma$-order) $\Gamma = 5$ will be used in the example instead of the more common choice $\Gamma = 4$ [29, 31]. The equivalent rectangular bandwidths (ERB) of the part-tones are chosen roughly in accordance to the psychoacoustically determined ones [28, 29]. For simplicity the bandwidths are not adapted within voiced segments. (The bandwidths are comparatively insensitive parameters of a decomposition which is aimed at a separation of acoustic modes, having typically a substantially smaller bandwidth, and which is not aimed at resynthesis of the speech signal.)

Excluding for a moment cases with subharmonic period number $m > 1$, the characteristic auto-phase-locking of voiced speech implies part-tones which have best filter frequencies being centred on integer (harmonic) multiples of the analysis window specific estimate of the frequency of the FD. The choice of audiological bandwidths for the part-tone decomposition has the effect that we can distinguish a lower range of harmonic numbers characterized by guaranteed single harmonic (resolved) part-tones and a range of potentially multiple harmonic (unresolved) part-tones. In the resolved part-tone range $1 \leq j \leq 6$ the harmonic order $h_j$ is identical to the part-tone index $j$. To avoid a substantial over-completeness (and *a priori* correlation between neighbouring part-tones) in the unresolved range $6 < j \leq N$, the set of part-tones is pruned according to the respective ERBs. A typical set of part-tones may have the harmonic orders $\{h_j\} = \{1, 2, ..., 6, 8, 10, 12, 15, ...\}$. In particular for speech segments, which correspond to nasals or vowels it is typical that some of the part-tones in the (a priori) unresolved range are also dominated by a single harmonic acoustic mode. The under-completeness of the part-tones in the (a priori) resolved range has a welcome noise suppression effect.

The all pole approximation of the gammatone filters has the advantage of a fast autoregressive algorithmic implementation [31]. For theoretical reasons we prefer its description in terms of a matrix recursion with a lower triangular matrix $L$ of dimension $\Gamma$ which plays the role of the cascade depth of the cascaded first order autoregressive filter,

$$L X_t = \lambda \exp(i\omega_t) X_{t-1} + e_1 S_t \quad \text{with} \quad (4a)$$

$$L = \begin{pmatrix} 1 & 0 & \cdots & \cdots & 0 \\ -1 & 1 & \ddots & & \vdots \\ 0 & -1 & \ddots & \ddots & \vdots \\ \vdots & \ddots & -1 & 1 & 0 \\ 0 & \cdots & 0 & -1 & 1 \end{pmatrix} \quad (4b)$$

and $\Gamma$-dimensional vectors $X_t = \{v_t, w_t, \cdots, z_t\}'$, $e_1 = \{1, 0, \cdots, 0\}'$ and $X_0 = 0$. The scalar $\lambda$ represents the damping factor of every first order autoregressive filter and is simply related to the ERB of the $\Gamma$-tone filter, $\lambda = \exp(-a_\Gamma ERB)$, the $\Gamma$-order dependent factor $a_\Gamma$ being given e.g. in [31]. The complex phase factor $\exp(i\omega_t)$ defines the instantaneous centre filter frequency $F_t = \omega_t / 2\pi$ being simply related to the instantaneous angular velocity $\omega_t$ and $S_t$ represents the input signal being sampled at discrete times $t$. The unusual feature is the time dependence of the angular velocity $\omega_t$ which will be specified later. The inverse of matrix $L$ is the lower triangular matrix with ones on and below the diagonal. It can be used to obtain $X_t$ as a power series of matrix $L^{-1}$,

$$X_t = \sum_{t'=0}^{t} \prod_{k=t'+1}^{t} \exp(i\omega_k) \lambda^{t-t'} L^{-(t+1-t')} e_1 S_{t'}. \quad (5)$$

The filter output is represented by the last component $z_t$ of vector $X_t$. Therefore we are exclusively interested in the matrix element in the lower left corner of any power of matrix $L^{-1}$. For the $(n+1)^{th}$ power this element can easily be obtained by complete induction as the ratio of three factorials $(n+\Gamma-1)!/(\Gamma-1)!n!$. The product of the complex phase factors in (5) can be expressed in terms of a sum of time dependent phase velocities $\omega_t$. Taking into account the additional dependence on the part-tone index $j$, the output of the bandpass filter of part-tone $j$ is thus obtained as

$$z_{j,t} = \sum_{t'=0}^{t} \exp\left(i \sum_{k=t'+1}^{t} \omega_{j,k}\right) \lambda_j^{t-t'} \frac{(t-t'+\Gamma-1)!}{(\Gamma-1)!(t-t')!} S_{t'}. \quad (6)$$

For $j = 1, ..., N$ the set of part-tones (6) can be interpreted as a highly over-sampled time-frequency decomposition of the speech signal $S_{t'}$, where the over-sampling is restricted to the time axis. The part-tone (6) is used to generate the part-tone phases (carrier phases)

$$\varphi_{j,t} = \arctan(im(z_{j,t})/re(z_{j,t})) \quad (7)$$

as well as the normalized part-tone phases $\varphi_{j,t}/h_j$ (in the frequency range of the pitch). If the trajectory (contour) of the centre filter frequency $\omega_{j,k}/2\pi$ is chosen as identical to the one of the instantaneous frequency $\omega'_k/2\pi$ of a constant amplitude input signal $S_{t'} = A\exp(i\sum_{k=0}^{t'}\omega'_k)$, the application of bandpass filter (6) generates the output

$$z_{j,t} = A \exp\left(i\sum_{k=0}^{t}\omega'_k\right) \sum_{t'=0}^{t} \lambda_j^{t-t'} \frac{(t-t'+\Gamma-1)!}{(\Gamma-1)!(t-t')!}. \quad (8a)$$

This filter output has the remarkable property that its instantaneous phase velocity is identical to the one of the input signal. For a given filter frequency contour, other input signals experience a damping due to interference of the phase factors. For a given input frequency contour, other filter frequency contours generate a phase distortion of the output. In the limit $t \to \infty$, the sum in equation (8a) represents an asymptotic gain factor. Being dependent on the bandwidth $\lambda_j$ and the $\Gamma$ order, the gain factors

$$g_{j,\Gamma} = \sum_{t'=0}^{\infty} \lambda_j^{t'} \frac{(t'+\Gamma-1)!}{(\Gamma-1)! \, t'!} \quad (8b)$$

can be used to obtain normalized part-tone amplitudes $a_{j,t} = |z_{j,t}|/g_{j,\Gamma}$. For more general voiced input signals the determination of filter frequency contours, which are identical to frequency contours of some underlying acoustic modes, represents a nontrivial problem.

Conventionally [24-27, 33] the adjustment of the filter frequency contours of part-tones (or "sinusoidal components") is achieved by introducing a short-time stationary (zero-chirp) subband decomposition which is densely sampled with respect to frequency and by determining for each point in time local maxima of the amplitudes of the subbands with respect to frequency. In a second step the maximizing frequencies of consecutive points in time are tested, whether they are suited to form continuous frequency contours. Suitable maxima are joined to form weakly non-stationary contours and part-tones. It is well known that the non-stationarity of natural voiced speech leads to frequent death and birth events of such contours, even within voiced segments [24-26].

The present approach is aimed at *self-consistent* centre filter frequency contours which are chosen as identical (or as consistent) to the frequency contours of the respective part-tones (outputs). The approach is based on the assumption that (sustained) voiced signals are composed of several part-tones which can iteratively be disclosed and confirmed to be self-consistent, when starting from appropriate contours of the centre filter frequency. More precisely, a part-tone reconstruction of a non-stationary acoustic object is self-consistent, if the centre filter-frequency contour(s) of a subset of the bandpass filters being used to generate the part-tone reconstruction is chosen as a stable invariant set of the iteration of two cascaded mappings, where the first mapping uses a set of filter-frequency contours to generate a set of part-tone phase velocity contours and the second mapping relates the same set of part-tone phase velocity contours to a set of filter-frequency contours which can be interpreted as an update of the mentioned ones. Whereas the first mapping is given by part-tone filter (6) and phase definition (7), the second mapping is chosen according to the acoustic properties of the assumed underlying physical system.

The second mapping includes in particular the time-scale separation step which generates smoothed filter-frequency contours. This implements the physical law *natura non facit saltus* and is suited to improve the convergence properties of the adaptation. Being inspired by equation (3b) the smoothing step might simply be chosen as a linear approximation of the trend of the filter-frequencies within each analysis window. However, due to the time reversal asymmetry of the $\Gamma$-tone filters, a negative chirp rate leads to a singularity of the instantaneous period length of the impulse response at finite times. This singularity can be avoided, if the time dependence of centre filter frequency $\omega_{j,t}/2\pi$ of part-tone $j$ is chosen separately depending on the sign of the (relative) chirp rate $c_j$. For negative chirp rate it is useful to assume alternatively a linear trend of the inverse of the respective centre filter frequency with a smooth transition at zero chirp rate

$$\omega_{j,k} = \begin{cases} \omega_{j,0}(1+c_j k) \\ \omega_{j,0}/(1-c_j k) \end{cases} \text{for} \begin{cases} c_j \geq 0 \\ c_j < 0 \end{cases}. \quad (9)$$

In case of a *single part-tone stable* non-stationary acoustic object the set of centre filter frequency contours and the set of part-tone phase velocity contours reduce each to a single contour. This case will be explained in more detail in the following section. In case of a *multi part-tone stable* non-stationary acoustic object it is useful to split up the mentioned second mapping into an intermediate mapping which generates the trajectory (contour) of the fundamental phase velocity $\omega_t$ as function of the set of part-tone phase velocity contours (and potentially also of the part-tone amplitudes), and into a third mapping $\omega_{j,k} = h_j \omega_k$, which represents the angular velocities $\omega_{j,k}$ of the filters as product of the part-tone index $j$ dependent phase normalization factor $h_j$ and of the time dependent fundamental phase velocity $\omega_k$. The third mapping expresses the nonlinear coupling of the underlying acoustic modes being described in the preceding section.

The phase normalization factors (winding numbers) $h_j$ are chosen either as given integer valued function of the part-tone index (see above) or as result of the cluster analysis. In the latter case the winding numbers may also assume rational values $n/m$ with the common denominator $m > 1$ representing the subharmonic period number. An example in place would be the set of winding numbers $\{h_j\} = \{1, 2, ..., 6, 15/2, 18/2, ...\}$. As an important additional self-consistency constraint of the reconstruction of the voiced excitation (primary response) according to equation (2), the subharmonic period number $m \geq 1$ has to be chosen identical to the one of the fundamental phase dependent coupling function $G(\psi)$ of equation (2).

The intermediate mapping incorporates the coincidence clustering step outlined in the preceding section. Due to the close connection to the acoustic correlate of virtual pitch perception, the details of the intermediate mapping should refer to the extended literature on related psychoacoustic experiments [26, 28]. In particular it is expected that the robustness of the reconstruction of the fundamental phase can be improved by making the intermediate mapping also dependent on part-tone amplitudes [27, 33].

The convergence properties of the self-consistent adaptation can further be improved by learning from hear physiology and psychoacoustics, in particular from the important role of the modulation amplitudes (envelopes) of the critical subbands [29]. As we will see in more detail it turns out to be advantageous, to base the reconstruction of the fundamental phase and in particular the cluster analysis not only on the described *carrier* phases but also on phases which are derived from the *envelopes* of part-tones. Being used exclusively in the range of unresolved harmonics, the envelope phases are determined e.g. as Hilbert phases of (appropriately scaled and smoothed) modulation amplitudes (envelopes) of part-tones.

To achieve a more uniform time evolution of the envelope phases and in agreement to well known results from hear-physiology and psycho-acoustics [28, 29], the (normalized) modulation amplitudes $a_{j,k} = |z_{j,k}| / g_{j,\Gamma}$ are submitted to a sublinear transformation (scaling) and smoothing prior to the determination of the Hilbert phases. The exponent $\nu$ of the sublinear scaling has been chosen in accordance to equation (1). In contrast to the carrier phases (which do nor need a correction due to their self-consistency as expressed in equation (8a)) the envelope phases need a group delay correction of the respective part-tones [31]. The part-tone index specific part of this correction has been derived from the maxima of the amplitude of the impulse responses of equation (6) [31]. The relative importance of the envelope phases is expected to increase, when the transmitter changes from a modal voice to a breathy one.

## 5. SELF-CONSISTENT PART-TONES OF A CHIRPED PHONATION PROCESS

To demonstrate the generation of self-consistent part-tones of a non-stationary voiced acoustic object and in particular to illustrate the convergence behaviour of the adaptation of their filter frequencies, several more or less drastic simplifications are introduced. Firstly the self-consistency concept is restricted to the simpler case of single part-tone stability and secondly the extraction of self-consistent part-tones is restricted to a sequence of synthetic glottal pulses chosen as a chirped sequence of constant amplitude saw teeth with a power spectrum, which is roughly similar to the one of the glottal excitation. Based on the deterministic skeleton of equation (2), the saw tooth shape of the pulses is described by the coupling function

$$G(\psi') = \min(\mod(\psi', 2\pi), s(2\pi - \mod(\psi', 2\pi))) \quad (10)$$

where $\psi'$ represents the phase of the glottal oscillator. The parameter s (chosen to be 6) determines the ratio of the modulus of the downhill slope of the glottal pulses to the uphill one. The chirp of the glottal oscillator is described by a time dependent phase velocity $\omega'(t') = \dot{\psi}'(t')$ which is chosen in analogy to equation (9), however, with potentially different chirp rate $c'$ and initial phase velocity $\omega'_0$

$$\omega'(t) = \begin{Bmatrix} \omega'_0 (1 + c't) \\ \omega'_0 / (1 - c't) \end{Bmatrix} \quad for \quad \begin{Bmatrix} c' \geq 0 \\ c' < 0 \end{Bmatrix}. \quad (11a)$$

In the more specific example the glottal chirp parameter $c'$ is chosen to generate a doubling of the frequency (or period length) after about 25 periods. The fundamental phase $\psi'$ is obtained by integration of $\omega'(t')$ as

$$\psi'(t) = \begin{Bmatrix} \omega'_0 (t + c' t^2 / 2) \\ -\omega'_0 / c' \ln(1 - c't) \end{Bmatrix} \quad for \quad \begin{Bmatrix} c' \geq 0 \\ c' < 0 \end{Bmatrix}. \quad (11b)$$

In the situation of signal analysis, appropriate contours of the centre filter frequency of the part-tone specific bandpass filters have to be obtained iteratively from the observed signal. As part of the time scale separation step of the second mapping of the last section, we assume that these contours can be described within the current (rectangular) window of analysis by a simple smooth function of time as indicated by equation (9). The single part-tone adaptation of the filter-frequency contour of the bandpass filter of part-tone $j$ can thus be reduced to an adaptation of parameters of equation (9). To reduce the dependence of the estimates on the size and position of the window of analysis (and/or to avoid the adaptation of the window length to the instantaneous period length), time scale separation ansatz (9) is extended by a $2\pi m$ periodic function $P_j(\varphi_{j,t}/h_j)$ of the respective normalized part-tone phase

$$\dot{\varphi}_{j,t}/h_j = \alpha_j t + P_j(\varphi_{j,t}/h_j) \quad (12a)$$
$$h_j/\dot{\varphi}_{j,t} = -\alpha_j t + P_j(\varphi_{j,t}/h_j). \quad (12b)$$

The $2\pi m$ periodic function $P_j(\varphi)$ accounts for the periodic oscillations of the phase velocity around the long term trend being generated by the characteristic auto phase-locking and is approximated by an appropriate finite order Fourier series. The Fourier coefficients as well as the trend parameter $\alpha_j$ are obtained by multiple linear regression.

Within a voiced segment of speech the adaptation of the parameters is performed sequentially for successive analysis windows. The initial value of the centre filter frequency of the current window is therefore typically given as result of the adaptation of the filter chirp of the preceding analysis window. Thus we treat the latter parameter as given ($\omega_{j,0} = h_j \omega_0'$) and concentrate on the convergence properties of the chirp parameter of the filter-frequency contour. The adaptation of a single parameter can be represented graphically. To explain the approach to self-consistency we use a graph, which shows the trend parameter $\alpha_j$ of equations (12a or 12b) for several part-tone indices $j$ as function of the common filter chirp rate $c$. To make figure 2 suited for the graphical analysis it gives the estimates of the relative trend $\alpha_j/(\omega_0' c')$ for the indices $j = 2, 4, 6, 9$ (corresponding to the sequence of the fixed points from bottom to top) as function of the relative filter chirp rate $c/c'$.

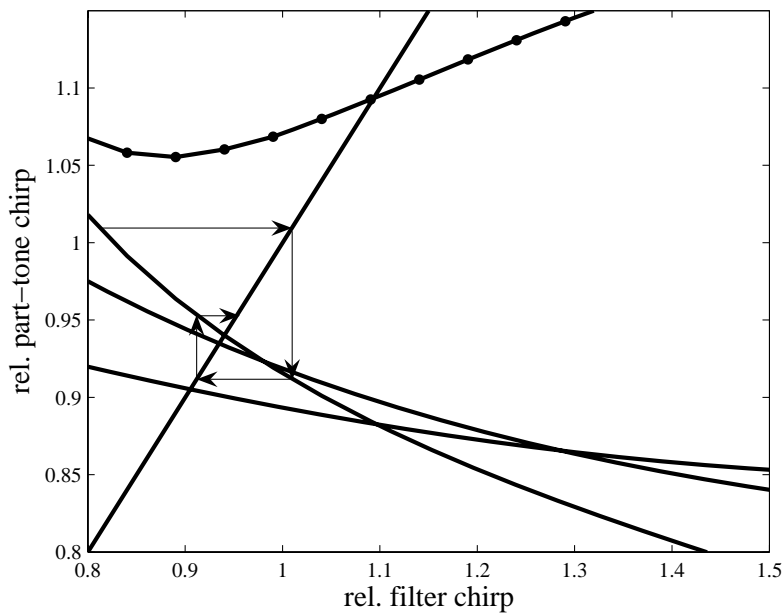

**Figure 2**: Estimated relative part-tone chirp rates as function of the relative chirp rate of the respective centre filter frequency, given for the envelope phase of part-tone 9 (circles, top) and the three carrier phases of part-tones 2, 4, and 6 (lines crossing the diagonal from bottom to top). All chirp rates are given relative to the chirp rate of the input sawtooth process defined in equations (10-11). The arrows and the diagonal of the first quadrant explain the algorithm, to determine the self-consistent centre filter frequencies.

The iterative adaptation of the chirp parameter of the filter-frequency contour can be read off from figure 2 by an iteration of two geometric steps: Project horizontally from one of the described curves to the diagonal of the first quadrant (which indicates the line where the fixed points of the iteration are situated) and project vertically down (or up) to the curve again. As can be seen from figure 2, the chirp parameters of all four part-tones have a stable fixed point (equilibrium) within a well extended basin of attraction of the chirp parameter which exceeds the shown interval of the abscissa. The fixed points (corresponding to the more general invariant sets of the preceding section) indicate the final error of the filter chirp which depends not only on the part-tone index but also on the size of the analysis window (which was chosen to have a length of about five periods of the glottal process). Due to the simple least squares regression of equations (12a,b), the modulus of the trend $\alpha_j$ is systematically underestimated.

The mutual phase locking of the self-consistently reconstructed part-tones is shown in figure 3. It demonstrates that the precision of reconstructed synchronization manifolds is hardly influenced by the deficient estimate of the filter chirp. The top row shows the perfect one dimensional synchronization manifolds in the state space projections spanned by the normalized *carrier* phases of the part-tones 5 and 6 (being used as ordinates) and the fundamental phase (being used as abscissa). The fundamental phase (with arbitrary initial phase) has been derived from the velocity of the carrier phase of part-tone 4. The bottom row shows the corresponding phase diagrams for the *envelope* phases of the part-tones 5 and 6. In all cases time invariant phase relations become visible. Due to the multiplicity of the definition of phases, the time invariance has to be valid also for relations between wrapped up phases. This implies that time invariant phase relations (so called circle maps) must have integer or rational valued average slopes (so called winding numbers).

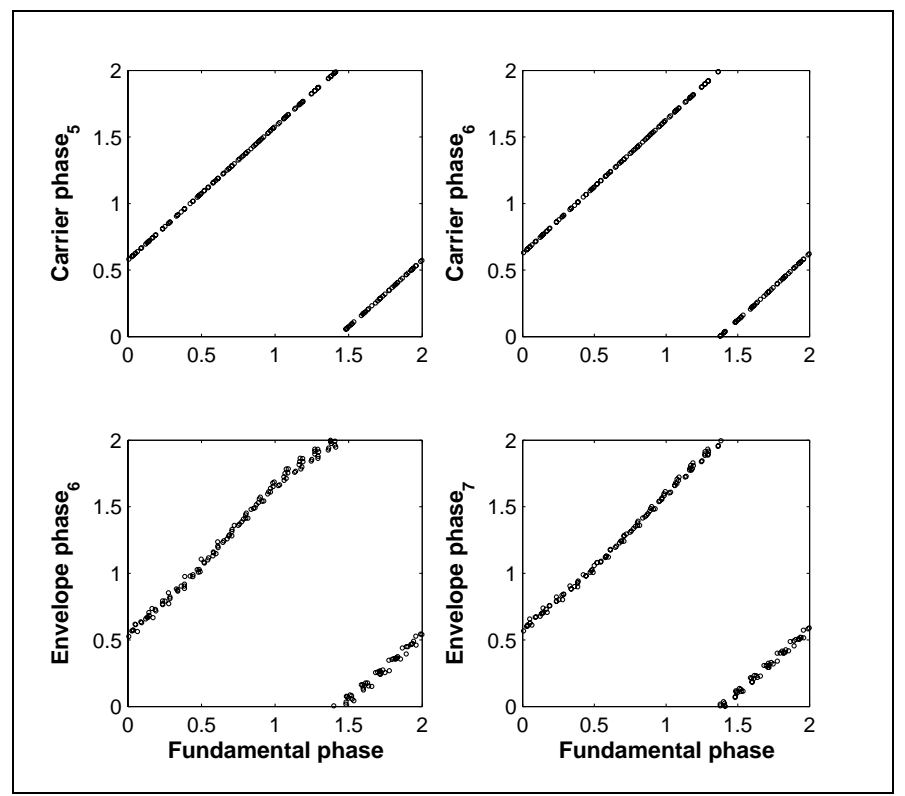

**Figure 3**: Circle maps (phase relations) of different part-tone phases which are suited to retrace the causal connection to the fundamental drive. Both carrier phases are normalized (divided by their part-tone specific integer winding number). Care has been taken that the wrapping of all phases happens simultaneously. All resulting phases are given in units of $\pi$.

Note that constant phase shifts of the part-tone phases due to finite (time invariant) group delays e.g. of the part-tone specific bandpass filter or of a stationary vocal tract filter are irrelevant at this stage of the analysis. In case of circle maps, which relate exclusively carrier phases (top row), the phase relations are precisely time invariant and result in precisely linear synchronization manifolds. In case of circle maps which relate mixed type phases (bottom row), the time invariance is achieved only approximately (by the described open loop group delay correction) and the circle map becomes curved, the shape being dependent on the described smoothing and sublinear scaling of the modulation amplitudes. When appropriately smoothed, the envelope part-tones are suited to be included into the cluster analysis of the phase velocity contours and potentially also to be used for a rough estimate of the phase velocity of the FD. They are, however, less suited for its precise reconstruction over an extended voiced speech segment.

## 6. Discussion

As has been pointed out by Terhardt [26, 27], human pitch perception can be trained to switch between analytic listening to a spectral pitch and synthetic listening to a virtual pitch. It is thus plausible to correlate the described single part-tone stable acoustic objects (with a macroscopic basin of attraction of the filter frequency contour or contour parameters) to outstanding part-tones, which are potentially perceived as spectral pitches by analytic listening [26, 27]. The number of stable invariant sets (fixed points) with a macroscopic basin of attraction depends in particular on the width of the power spectrum of the voiced signal. In the example of the last section a strong asymmetry of the sawteeth ( $s \gg 1$ in equation 10) favors the stability of higher order fixed points.

From psychoacoustic experiments it is also known that virtual pitch is a more universal and robust percept than spectral pitch [4, 26]. Based on the *a priori* assumption that the signal is generated by a voice production system, which generates several phase locked higher frequency acoustic modes, the observed (carrier or envelope) phase velocity of one part-tone might be used to adjust the centre filter frequency of other part-tones. This opens the possibility to use the more robust multi part-tone adaptation strategy which can be expected to converge even in cases with no single part-tone stability. The sensation of virtual pitch is thus expected to be correlated to the existence of a multi part-tone stable acoustic object with a macroscopic basin of attraction of the filter frequency contour (parameters). The graphical analysis of figure 2 gives several hints to the design of an optimized multi part-tone stable self-consistency generating decoder. Envelope phases (circles) are less suited for an accurate adaptation of the filter frequency contours than the carrier phases. Carrier phases of the higher harmonic part-tones are better suited to obtain precise fixed points than the low order carrier phases. However, the size of the basin of attraction of the initial filter frequency ( $\omega_{j,0}/2\pi$ ) becomes necessarily smaller for higher harmonic part-tones.

The design principle of a robust, fast and precise algorithm to extract a self-consistent fundamental phase velocity contour from the speech signal can thus be summarized as follows: Use the envelope phases and/or low order carrier phases to find filter frequency contour parameters which lie in the basin of attraction of stable fixed points of higher order resolved part-tones - preferentially of those part-tones which are *(n:n')* phase locked to the carrier of other part-tones. Such part-tones are well suited to adapt the centre filter frequencies and to reconstruct the fundamental phase. In fact it turns out that this design principle is suited to explain several well known properties of virtual pitch perception including the fact that pitch perception prefers (jumps on) the resolved part-tones, in particular the harmonic part-tones with $h_j \geq 3$, if they are (consistently) available as part-tones of the acoustic signal [26, 27, 28]. The coincidence between the theoretically derived properties of an efficient transmission protocol for a multi part-tone stable acoustic object and the empirically observed properties of virtual pitch perception gives additional support to the interpretation of the instantaneous frequency of the FD as acoustic correlate of pitch perception. The coincidence is also suited to underline the central role of the fundamental phase for decoding voiced speech.

The example of the last section is characterized by more than 15 pairs of mutually synchronous carrier phases. In that case the highest harmonic order with a phase locked part-tone is only limited by the bandwidth of the part-tone filters. In the case of voiced speech signals the number of self-consistent and/or phase locked part-tones is expected to be additionally limited by characteristic properties of the speech signal. Part-tones, which are not suited to confirm the topological equivalence of the FD, are also defined uniquely by filter frequency contours which represent integer or rational multiples of the self-consistent fundamental frequency.

Being a well adapted transmitter for noisy communication channels, the production system of voiced speech can be expected to operate as a DR system with a well balanced coupling between the hidden drive and the observable response, where the balanced coupling is strong (direct) enough to enable a robust topologically equivalent reconstruction of the hidden drive and sufficiently weak (indirect) to open the option of a qualitatively rich spectrum of different (finite dimensional) response dynamics. The resulting voiced speech is thus well distinguishable from the (infinite dimensional) stochastic noise of the communication channel and from other voices being driven by different (asynchronous glottal) master oscillators. The speaker specific differentiation of the dynamics (e.g. in terms of part-tone specific winding numbers) is expected to be extremely stable (nearly lifelong), whereas the phoneme specific differentiation is expected to show a fast variation (hopping) in comparison to typical disturbances of the acoustic communication channel like movements of the speaker. The assumed long time stability of the part-tone specific winding numbers can only be understood as result of a closed loop neural control. Due to the described balance of the coupling strength between the glottal master oscillator and the part-tones of the speech signal, the reconstruction of the fundamental phase of colloquial speech might not be a simple task. However, the astonishing ubiquity, robustness and precision of human pitch perception point into the direction that this task can finally be achieved.

The coherent reconstruction of the fundamental phase of a complete voiced speech segment as well as the combined event and wave type description of the speech signal are particularly suited to analyse voiced consonants. In conventional speech analysis, short term stationary deterministic and/or stochastic process models of the speech signal represent the highest *acoustic* level description of speech, being followed in hierarchy by a *symbolic* (finite state Markov chain) description of the non-stationary properties of speech. Though the introduction of the non-stationary self-consistent acoustic objects has nearly doubled the length of the basic window of analysis, there are good reasons to assume that gliding coordination of part-tone phases plays an important role in speech communication and that the acoustic process description of voiced speech should be extended further, up to the level of uninterrupted voiced segments of syllables. The analysis of the gliding coordination of the part-tone phases opens the option to transport the time (fundamental phase) stamp of the glottal closure event through a complete voiced speech segment. The knowledge about the time stamp can be used with advantage for the analysis of the long range correlation of the responses of nasals and other sustainable voiced consonants [8, 14].

The described long range correlation in voiced segments of speech brings new light to the answer of an old question in phonetics, whether the phones or the syllables are the atomic symbols of speech communication. In this context it is important to remind that many phones are so far defined exclusively by human perception of minimal differences between syllables. The present study has outlined an additional plausible cue of self-consistent acoustic objects (the fundamental phase shift of the glottal closure event) which can only be extracted from an acoustic process description of a sequence of phones. On the other hand some of the classical symbols of phonetics (e.g. stop consonants) can be interpreted as a characteristic sequence of at least two non-stationary acoustic objects which partially do themselves represent super-positions of several or many elementary time-frequency atoms. In contrast to the phonetically relevant components of speech the elementary "atomic" status of the latter components is guaranteed by a well defined physical principle [30] proposed by Heisenberg.

## 7. CONCLUSION

A transmission protocol of non-stationary self-consistent (voiced) acoustic objects is outlined, which are described as stationary response of a non-stationary fundamental drive (FD) and which can self-consistently be decomposed into non-stationary part-tones. Self-consistent part-tones are characterized by phase velocities which are consistent with the centre filter frequencies being used to generate the (complex) part-tones. The second property of the self-consistent acoustic objects qualifies them as most elementary objects of a voice transmission protocol which is centred on a time scale separation with a precise and robust decoding option. It is hypothesized that the self-consistent decomposition of speech segments, which are suited to transmit voiced continuants, leads to a subset of part-tones which shows generalized synchronization of their phases. The iterative identification of multi part-tone stable voiced acoustic objects relies on and enables a high precision reconstruction of a fundamental phase which can be confirmed as phase of a topologically equivalent image of a glottal master oscillator on the transmitter side. As topologically equivalent image on the receiver side, the self-consistent FD represents the long time scale part of the basic time scale separation of human acoustic perception. The latter separates the timbre or phone specific part of a sound from pitch and loudness or intonation and prosody. The self-consistent reconstruction of the FD avoids the assumption of a frequency gap being necessary to justify the assumption of a stationary or periodic voice source. The latter assumption is conventionally used to estimate phone specific cues and/or the pitch contour.

The high frequency part of the time scale separation of a speech signal is conventionally described by the source-filter model. The voice source (excitation of the vocal tract filter) is conventionally determined by applying an inverse filter to the speech signal. As alternative, the broadband voiced excitation is determined as synchronized primary response of the band-limited FD. The part-tone specific (periodic) fundamental phase dependent coupling functions of the primary response are expected to show a wealth of qualitative features which are related to periodicities and time shifts of underlying acoustic modes on the transmitter side. These quantities represent topological invariants (like the better known decay rates of vocal tract responses), which are minimally disturbed by changes of distance and direction of the acoustic communication channel.

The question in how far the transmission protocol of non-stationary self-consistent acoustic objects is actually used for the distinction of speakers and phonemes in human speech communication, has to remain largely open at the moment. The surprisingly subtle transmission protocol may, however, have emerged as result of a combined phylogenetic and ontogenetic co-evolution of the sound production system and the auditive pathway in a highly variable acoustic environment which is strongly influenced by vocalizations of the contemporary members of the own species. The efferent enervation of the outer hair cells of the cochlea confirms that there is no obvious physiological contradiction to a centrally closed loop controlled focussing on voice specific part-tones in the cochlea.

*Acknowledgements:* The author would like to thank M. Kob, B. Kröger, C. Neuschaefer-Rube, R. Schlüter, Aachen, J. Schoentgen, Brussels, A. Lacroix, K. Schnell, Frankfurt, P. Grassberger, H. Halling, M. Schiek, P. Tass, Jülich, N. Stollenwerk, Lisboa, V. Hohmann, B. Kollmeier, J. Nix, Oldenburg, and J. Rouat, Québec for helpful discussions and/or comments.